\documentstyle[aps,pre,multicol,psfig,rotate]{revtex}
\begin{document}
\draft 
\tighten

\title{Largest Lyapunov Exponent for Many Particle Systems at Low 
Densities} 
\author{R. van Zon[*],  H.\ van Beijeren[*] \and Ch. Dellago[**]}
\address{* Institute for Theoretical Physics, University of Utrecht,
Postbus 80006, 3508 TA, Utrecht, The Netherlands, \\ $**$  
Department of Chemistry, University
of California, Berkeley, CA, 94720, U.S.A.}
\date{\today}
\maketitle

\begin{abstract}

The largest Lyapunov exponent $\lambda^+$ for a dilute gas with
short range interactions in equilibrium is studied by a mapping
to a clock model, in which every particle carries a watch, with a
discrete time that is advanced at collisions.  This model has a
propagating front solution with a speed that determines
$\lambda^+$, for which we find a density dependence as predicted
by Krylov, but with a larger prefactor. Simulations for the clock
model and for hard sphere and hard disk systems confirm these
results and are in excellent mutual agreement. They show a slow
convergence of $\lambda^+$ with increasing particle number, in
good agreement with a prediction by Brunet and Derrida.

\end{abstract}


\newcommand{\vect}[1]{\vec{#1}}
\newcommand{\vv}{\vect{v}}
\newcommand{\vrr}{\vect{r}}
\newcommand{\vecd}[1]{\delta\vec{#1}}
\newcommand{\vdv}{\vecd{v}}
\newcommand{\vdr}{\vecd{r}}
\newcommand{\matr}[1]{\mbox{\boldmath $#1$}}
\newcommand{\eq}[1]{Eq.~(\ref{eq:#1})}
\newcommand{\eql}[1]{\label{eq:#1}\label{#1}}
\newcommand{\ddh}[2]{\frac{d #1}{d #2}}
\newcommand{\order}{{\cal O}}
\newcommand{\vhat}[1]{\hat{#1}}

\begin{multicols}{2} \narrowtext

  Recently, there has been great interest in the relationship
between statistical mechanics and the theory of dynamical
systems\cite{Evans,Mareschal,Dorfman2}. Calculating dynamical
properties such as Lyapunov exponents for statistical mechanical
systems usually requires numerical simulations. For the Lorentz
gas however, Dorfman, Van Beijeren and others have obtained
analytical expressions for the Lyapunov spectrum and
Kolmogorov-Sinai entropy at low densities, both in equilibrium
and for the field-driven case\cite{Dorfman2,Beijeren}.

  In this paper we present an analytic calculation of the largest
Lyapunov exponent in the low density limit for a gas at
equilibrium consisting of particles with short range
interactions.  Our method is based on arguments from kinetic
theory and similar in spirit to the method of
Refs.~\cite{Dorfman2,Beijeren}.  We compare our results to those
from computer simulations on hard disk and hard sphere systems
and pay special attention to the dependence of the largest
Lyapunov exponent on the total number of particles. 

  We consider a gas consisting of $N$ atoms of diameter $\sigma$,
defined as the (strictly finite) range of interaction, and mass
$m$ in $d$ dimensions, in a volume $V$.  The reduced density
$\tilde{n}$ is defined as $ N\sigma^d/V$ and will serve as a
small parameter. To calculate the largest Lyapunov exponent we
follow two nearby trajectories in phase space. For the first one,
the reference trajectory, the positions and velocities of the
particles are denoted by $(\vrr_i,\vv_i)$. In the second
trajectory they are denoted by $(\vrr_i+\vdr_i,\vv_i+\vdv_i)$.
The {\em deviations} $(\vdr_i,\vdv_i)$ will be taken to be
infinitesimally small. For a chaotic system, they will grow
exponentially with time at a rate equal to the largest Lyapunov
exponent $\lambda^+$. Since the whole vector $(\vdr_i,\vdv_i)$ in
phase space grows exponentially, so will a generic projection,
hence one has
\begin{equation}
	\lambda^+=\lim_{t\rightarrow\infty} \frac{1}{2t} 
	\ln
	\left[ 
	 \frac{\sum_{i=1}^{N}\|\vecd{v}_i(t)\|^2 }
	{\sum_{i=1}^{N}\|\vecd{v}_i(0)\|^2 } \right].
 \label{lyap}
\end{equation}
Therefore, in order to calculate $\lambda^+$ one has to find out
how $\vecd{v}_i(t)$ typically increases with time. We will first
illustrate this on the somewhat simpler case of the random
Lorentz gas, consisting of a single light particle moving among a
random array of fixed scatterers interacting with the light
particle through a spherically symmetric potential. Between
collisions the velocity deviation does not change and the
position deviation changes according to
\begin{equation}
        \vdr (t) =\vdr(t_0) + (t-t_0)\vdv(t_0) .
 \label{deltar}
\end{equation}
In a collision the velocity changes from $\vv$ to $\vv'$ given by
\begin{equation}
   \vv' = \vv - 2 (\vhat{n}\cdot\vv){\vhat{n}}\equiv 
   \matr{M}_{\vhat{n}} \vv.
 \label{v'}
\end{equation}
$\vhat{n}$ denotes the unit vector in the direction from the
center of the scatterer to the point of closest approach.  The
change of $\vdv$ in a collision is obtained from Eq.~(\ref{v'})
by expanding both $\vv +\vdv$ and $\hat{n}+\delta\hat{n}$ to
linear order in the deviations. The difference in impact times
for the two nearby trajectories leads to a shift in $\vdr$. 
Since deviations follow linearized dynamics one always has
\begin{equation}
   \left(\begin{array}{c} \vdr' \\ \vdv' 
   \end{array}\right) =
   \left(\begin{array}{ccc} \matr{A} &&2\matr{P} \\ -2\matr{Q} && 
   \matr{B} \end{array}\right)
   \left(\begin{array}{c} \vdr \\ \vdv 
   \end{array}\right) .
 \eql{transdv}
\label{matrix}
\end{equation}
For hard sphere scatterers with radius $\sigma$ it turns out
that, in any number of dimensions,
$\matr{A}=\matr{B}=\matr{M}_{\vhat{n}}$, $\matr{P}=0$ and
\newcommand{\identity}{\matr{1}}
\begin{equation}
  \matr{Q} =
  [\sigma(\vhat{n}\cdot\vv)]^{-1}
  \left[
          (\vhat{n}\cdot\vv)\identity + \vhat{n}\vv
       \right]\cdot\left[
          (\vhat{n}\cdot\vv)\identity - \vv \vhat{n}
       \right],
 \eql{Q}
\end{equation}
with $\identity$ the identity matrix.  A derivation of these
results in two dimensions can be found in \cite{Dellago}. From
the above equations we infer that at low density, just after the
$k$'th collision, with $k$ very large, $\vdv$ and $\vdr$ will
typically have increased to
\begin{equation}
   \vdv'(t_k)\approx v\left( \alpha /{\tilde{n}}\right)^k \; ; \;
   \vdr'(t_k)\approx \sigma \left( \alpha /{\tilde{n}}\right)^k ,
 \label{deltak}
\end{equation}
with $v$ the speed of the light particle, and $\alpha$ a constant
of order unity.  This follows from an inductive argument: suppose
Eq.~(\ref{deltak}) is valid after the $k$'th collision, then
according to Eqs.~(\ref{deltar}) and (\ref{deltak}) one has
$\vdr(t_{k+1}) \approx \vdr'(t_{k}) +t_{mf}\vdv'(t_k) \approx
\sigma (\alpha/\tilde{n})^{k+1}$, where we replaced $t_{k+1} -
t_k$ by its average value, the mean free time $t_{mf}$. In the
last approximate equality we neglected $\vdr'(t_k)$ since it is
one order of $\tilde{n}$ smaller than $t_{mf}\vdv'(t_k)$. 
According to \eq{transdv} and (\ref{deltak}), after the $(k+1)$th
collision $\vdv'(t_{k+1}) = \vdv(t_k) - 2 \matr{Q} \vdr(t_{k+1}) 
\approx v (\alpha/\tilde{n})^{k+1}$, where we neglected
$\vdv(t_k)$ since it is one order of $\tilde{n}$ smaller than the
second term, and used that the typical size of the matrix
elements of $\matr{Q}$ is $(v/\sigma)$, as is explicitly seen in
\eq{Q}.  Now, because $t \approx k t_{mf}\equiv k/\nu$, with
$\nu$ the single particle collision frequency, it follows from
Eqs.~(\ref{lyap})  and (\ref{deltak}) that the Lyapunov exponent
is
\[
   \lambda^{+} = -\nu \ln \tilde{n} + \nu \ln \alpha ,
\]
with $\alpha$ to be determined by an averaging procedure over
free flight times and collision dynamics. 
This estimate was already obtainted by Krylov\cite{Krylov}.
Notice that the value
of $\alpha$ is not important for the dominant first term in
$\lambda^{+}$.  

  These considerations can be generalized to systems of identical
moving particles by noting that in a collision, say between
particles 1 and 2, Eqs.~(\ref{v'}-\ref{Q}) still are applicable
to the relative velocity, $\vv=\vv_1-\vv_2$, the relative
velocity deviation, $\vdv=\vdv_1-\vdv_2$ and the relative
position deviation $\vdr=\vdr_1-\vdr_2$.  In addition one needs
the corresponding relations for the center of mass coordinates
$\vec{V}=(\vv_1 +\vv_2)/2$ and $\vec{R}=(\vec{r_1}+\vec{r_2})/2$,
which are
\begin{equation}
   \vec{V'}=\vec{V} \; ; \; \delta\vec{V'}=\delta\vec{V} \;  ; \;
   \delta\vec{R'}=\delta\vec{R}.
 \label{cm}
\end{equation} 
Assume now that the deviations for particles 1 and 2 just after
their last collisions before the present one were of the form
(\ref{deltak}) with exponents $k_1$ respectively $k_2$, and $v$
the mean relative velocity. By similar reasoning as for the
Lorentz gas it follows that just before collision $\vdv$ and
$\delta\vec{V}$ both are of order 
$(\alpha/\tilde{n})^{max(k_1,k_2)}$ whereas $\delta\vec{r}$ and
$\delta\vec{R}$ are of order
$(\alpha/\tilde{n})^{max(k_1,k_2)+1}$. As a consequence of
(\ref{matrix}) and (\ref{cm}) right after the collision $\vdr'_i$
and $\vdv'_i$ ($i=1,2$) will then also be of order
$(\alpha/\tilde{n})^{max(k_1,k_2)+1}$.  So on average
$\ln|\vdv_i|$ also increases by units of $\ln (\alpha/\tilde{n})$
at collisions, but in contrast to the Lorentz gas this increase
may involve several of these units, in case the other particle
involved in the collision has a higher $k$-value. 

  The values of $\ln \alpha$ in an actual realization of the
dynamics will fluctuate strongly from collision to collision. 
However, their distribution becomes independent of
density and increasingly narrow relative to $\ln (1/\tilde{n})$
as density gets closer to zero. Therefore the essence of the
dynamics determining the largest Lyapunov exponent is captured in
the following simple clock model:  Think of each particle $i$ as
carrying a watch, whose clock value is $k_i$. When two particles
collide, they synchronize their watches to the larger of the two
clock values, and advance them by one unit. The largest Lyapunov
exponent will be determined by the speed $w$ by which the watches
run on average and will be of the form
\begin{equation}
   \lambda^{+} = w(-\nu \ln\tilde{n} + \nu \ln\alpha ).
 \label{lle}
\end{equation}  
The synchronization of the $k$-values prohibits a direct
identification with the number of collisions like we could do in
the Lorentz gas. 

\begin{figure}[p]
	\centerline{\psfig{figure=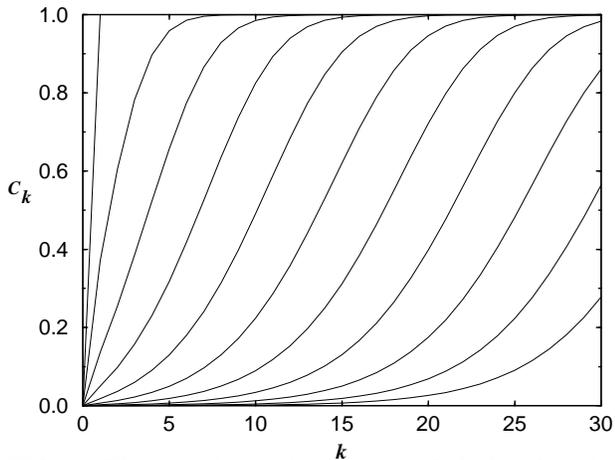,width=8cm,angle=-90}}
	\caption{The cumulative distribution of clock values $k$ at 
          times $\tau=0$, $\tau=1$, $\tau=2\ldots\tau=10$ (from left
          to right).}
 \label{fig:1} 
\end{figure}

  We will use a mean field approach to calculate the clock speed
$w$. We denote the number of particles that have a given clock
value $k$ by $N_k$ and assume that they are distributed uniformly
in $V$. In collisions involving particles with clock value $k$,
$N_k$ decreases.  It is increased by two in collisions in which
the largest incoming $k_i$ was $k-1$.  So the rate equations for
the $N_k$ become
\[
   \ddh{N_k}{t} = - \sum_{\stackrel{\scriptstyle l=-\infty}{l\not = 
   k}}^{\infty} R_{(k,l)} 
                  - 2 R_{(k,k)}
                  + 2 \sum_{l=-\infty}^{k-1} R_{(k-1,l)}.
\]
$R_{(k,l)}$ are the rates by which collisions between $k$ and $l$
take place. We use a Sto{\ss}zahlansatz: the rate of collisions
between particles with clock values $k$ and $l$ is proportional
to $N_{k}N_{l}/N^2$. Since all rates are also proportional to
$\nu$, we will express time in units of the mean free time:
$\tau=\nu t$. We use the fractions $f_k=N_k/N$ to eliminate the
$N$ dependence: 
\[
   \ddh{f_k}{\tau} = -f_k + 2f_{k-1}\sum_{l=-\infty}^{k-2} f_l + f_{k-1}^2.
\]
For the cumulatives $C_k=\sum_{i=-\infty}^{k} f_i$ this reduces to
\begin{equation}
   \ddh{C_k}{\tau} + C_k = C_{k-1}^2 .
 \eql{C}
\end{equation}
The solution is given by the recursion relation
\[
   C_k(\tau) = e^{-\tau} C_k(0) + \int_{0}^{\tau} e^{\tau'-\tau} 
   C_{k-1}^2(\tau') d\tau'. 
\]
If $C_k$ is zero at $\tau=0$ it remains zero. Thus the starting
point of this recursion is the smallest $k$ for which $C_k(0)\neq
0$.  Inductively we see that all $C_k$ are polynomials in
$e^{-\tau}$, of which the degree grows exponentially with $k$. 
We calculated these polynomials with initial conditions
corresponding to $f_k(\tau=0)=\delta_{k1}$. The exponentially
growing degree of the polynomials enables only a limited number
of $C_k$ to be computed, even on a computer.  The results up to
$k=30$ at several time values are shown in Fig.~\ref{fig:1}.  The
initial distribution broadens and moves to the right. We expect
the distribution to asymptotically become a front propagating at
a constant speed $w$. Then we have
\begin{eqnarray*}
   \sum_{i=1}^{N} \|\vecd{v}_i(t)\|^2  &=&
   \sum_{k=-\infty}^{\infty} f_k(\tau)\;v^2 e^{-2k
   \ln(\tilde{n}/\alpha)} \\
   &\simeq&\sum_{k=-\infty}^{\infty} f_{k}(0)\;v^2 
   e^{-2(k-w\tau)\ln(\tilde{n}/\alpha)} 
\end{eqnarray*}

\noindent
This result should be proportional to $e^{2t\lambda^{+}}$ so one
indeed recovers Eq.~(\ref{lle}).  It agrees with Krylov's
conjecture\cite{Krylov}, except for the appearance of $w$. The
result of Stoddard and Ford \cite{StoddardFord} is of the same
form when expanded in $\tilde{n}$. 

The speed $w$ should be independent of the initial conditions.
Physical initial conditions will have finite support, because
there is always a particle with smallest $k$ and a particle with
highest $k$.  Then $C_k(0)$ is bounded by two step functions.
Using \eq{C} one can show that they will remain bounded by the
solutions corresponding to these two initial conditions. If these
tend to some uniformly moving solution with speed $w$, so will
the real system.  Thus $w$ is unique for this set of initial
conditions. 

We put the propagating front Ansatz $C_{k}(\tau) = F(k-w\tau)$
into \eq{C} to obtain a differential-difference
equation\cite{Bellman} for the shape of the cumulatives:
\begin{equation}
   -w \ddh{F}{x}(x) + F(x) = F^2(x-1) .
 \eql{dde}
\end{equation}
$F$ has to be monotonically increasing, tending to $0$ as
$x\rightarrow -\infty$ and to $1$ as $x\rightarrow\infty$. This
means that $F=0$ has to be unstable and $F=1$ has to be stable. 
It is easy to see that these are fixed points of \eq{dde}. Their
stability is determined by linearized equations. The behavior
around a fixed point is always exponential: $F(x)=\sum_j p_j
e^{s_j x}$, in which the $s_j$ are roots of the so-called
characteristic equation and $p_j$ are polynomials in $x$ of
degree less than the multiplicity of root $s_j$ \cite{Bellman}.
For an unstable fixed point some of the $s_j$ have positive real
parts. Around $F=0$, this is true if $w>0$. For a stable fixed
point, the term with least negative $s_j$, let's call this
\newcommand{\leastneg}{\gamma} $-\leastneg$, will dominate the
large $x$ behavior. If $\leastneg$ were complex, we would see
oscillatory behavior, violating monotonicity, so $\leastneg$ has
to be real. Inserting the asymptotic behavior
$F(x)=1-\exp(-\leastneg x)$ into \eq{dde} and neglecting
quadratic terms, produces the characteristic equation $\leastneg
w+1-2e^{\leastneg} =0$.  This gives a relation between $w$ and
$\leastneg$: $w(\leastneg)= (2e^{\leastneg}-1)/\leastneg$. It
turns out that there is a minimum $w$ for positive real values of
$\leastneg$ which can be expressed in terms of Lambert's $W$
function:
\[
   w = -1/W(-1/2e) = 4.31107..
\]
This value is in accordance with estimates from Fig.~\ref{fig:1}. 
Solutions with initial conditions with finite support select this
minimum speed.  The same kind of velocity selection occurs in
other systems, for a number of which it has been
proved\cite{Benguria}. 

\begin{figure}[p]
   \centerline{\psfig{figure=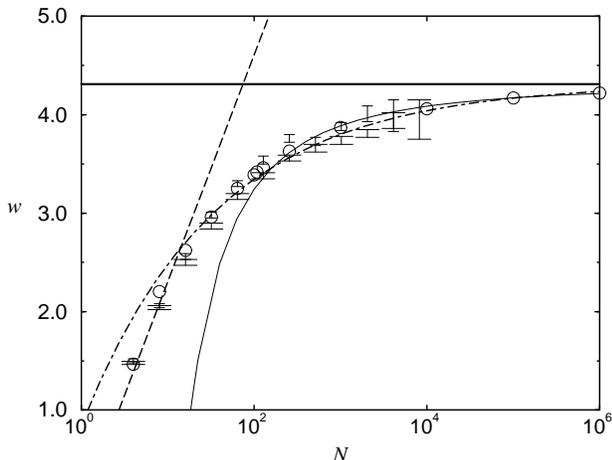,width=8cm,angle=-90}}
   \caption{Number dependent coefficient $w$. The circles are
    clock model results. The bars are new molecular simulation 
    results: wide error bars for hard spheres, narrow error bars
    for hard disks. The dashed line is Stoddard and Ford's $\ln N$
    prediction. The dashed-dotted curve is a fit of the mean
    field results to the algebraic expression (\ref{curve}).
    The thick line gives the analytic result for 
    $N\rightarrow\infty$. The solid curve is the prediction of
    \eq{wNprediction}.} 
 \label{fig:Ndep} 
\end{figure}

  We compared the result $w=4.311..$ with those from simulations
done by Dellago, Posch and Hoover\cite{Dellago}, with 64 hard
disks. They made a fit of the largest Lyapunov exponent to $a\;
n\ln(n/b)$ indicating a value of $w\approx 3.3$. The difference
to our value turns out to be due to large finite $N$ effects.
First we'll show this numerically.  We take $N$ watches and give
them some initial $k$ values. In each time step we pick two
watches at random and advance their $k$-values according to the
rules of the clock model.  We compute the average growth of $k$
per watch per time step, and find $w_N$.  We did this for numbers
of watches ranging from 4 to $2^{19}$. In Fig.~\ref{fig:Ndep} the
results were fitted to an algebraic curve \begin{equation}
w_N=4.311-A\,N^{-B}.  \label{curve} \end{equation} A good fit,
except for the smallest values of $N$, was obtained by choosing
$B=0.277$ and $A=3.466$.  No good fit could be obtained for $B=1$
or on replacing the algebraic $N$-dependence by an exponential
one. The value $B=0.277$ is in reasonable agreement with the
exponent of $-1/3$ obtained by Dellago and Posch\cite{Dellago3}. 
In order to come to a better comparison between clock model
predictions for $w_N$ and simulation results on actual dilute gas
models we performed a number of new simulations using the same
methods as in Refs. \cite{Dellago,Dellago3}, both on hard disk
and hard sphere systems for different particle numbers at a
number of low densities, $\tilde{n}=10^{-5}$, $10^{-4}$,
$10^{-3}$ and $10^{-2}$.  For each $N$ the results for the
largest Lyapunov exponent were fitted to $-w\nu \ln
(\tilde{n}/\alpha)$.  The results for $w$ are also plotted in
Fig.~\ref {fig:Ndep}.  One sees that the results of the clock
model are in excellent agreement with those of the simulations. 

  Stoddard and Ford\cite{StoddardFord} used crude arguments to
get $w_N=\ln N$. This relation is also plotted in Fig.
\ref{fig:Ndep}. One sees that this only gives a good fit for very
small $N$.  In simulations of 100 particles with a cut-off
Lennard-Jones interparticle potential, Stoddard and Ford found
agreement with their predicted value, which lies somewhat above
our asymptotic value of 4.311 and much more above the simulation
results for 100 particles, both in the clock model and for hard
spheres and disks. Stoddard and Ford acknowledged that the error
increases as $\tilde{n}$ gets smaller and say that their
simulation results for low $\tilde{n}$ should not be expected to
fit their theory.  But the $\tilde{n}$ at which the error becomes
too big, is not sharply defined.  If one takes it low enough,
the data support their prediction, but if one takes it a little
higher the results come more in line with those from the clock
model simulations. 

  In a recent paper \cite{Brunet}, Brunet and Derrida present a
way to compute the $N$ dependence of the velocity in a similar
model by treating it as a discretization effect.  In our case,
there are at least two particles with highest $k$ in any
realization.  This means we have to take $\epsilon=2/N$ in
equation (7) in \cite{Brunet}.  Inserting our expression for
$w(\leastneg)$, we find
\begin{equation}
   w_N	= w - \frac{(w-1) \pi^2}{2 \ln^2 (N/2)},
 \eql{wNprediction}
\end{equation}
We plotted this prediction also in Fig.~\ref{fig:Ndep}. The
agreement is good for $N>100$.

  In the work by Searles et al\cite{Searles} a weak but
persistent increase in $\lambda^+$ with $N$ was interpreted as a
sign of a logarithmic divergence.  It was argued that the data
were not consistent with a $1/N$-approach to a constant value and
a plot of $ \lambda^+$ versus $\ln N$ looks quite linear over an
appreciable range. However, Dellago and Posch\cite{Dellago3} in
their simulations on dense hard sphere systems did not observe
such a divergence and in fact it looks like the results of
Searles et al are entirely consistent with the type of behavior
predicted by Brunet and Derrida. The mean field analysis given
here is not decisive though, since it completely ignores all
effects of local density and temperature fluctuations.

  We conclude by stressing that the first term of the density
expansion of the largest Lyapunov exponent of a dilute gas that
was calculated in this letter, is universal for systems where the
interaction is sufficiently short ranged, i.e. it strictly
vanishes beyond its range $\sigma$, or perhaps may be allowed to
vanish exponentially.  This letter shows that the calculation of
dynamical properties of many particle systems is feasible and
that the calculation of the largest Lyapunov exponent in dilute
gases requires the solution of a nonlinear front propagation
equation. The method will be extended in future work to get the
$\order(\tilde{n})$ term of the Lyapunov exponent. This term will
depend on the details of the interaction at a collision, and is
of considerable physical interest. 

  We thank professors J.\ R.\ Dorfman and H.\ A.\ Posch for
valuable discussions, professor Th.\ W.\ Ruijgrok for pointing
out useful references and professor A.\ Zangwill for providing us
with a copy of the doctoral thesis of Stoddard. The work reported
here was supported by FOM, SMC and by the NWO Priority Program
Non-Linear Systems, which are financially supported by the
"Nederlandse Organisatie voor Wetenschappelijk Onderzoek (NWO)".
The authors acknowledge the hospitality and support of the
Institute for Physical Science and Technology at the University
of Maryland during the first stages of their research.

\end{multicols}

\end{document}